\documentstyle[aps,prl,epsf,twocolumn]{revtex}
\begin{document}
\draft 
\title{
{\rm Physical Review Letters}\hfill {\sl Version of \today}\\~~\\ 
Extended Self-Similarity in Turbulent Systems:
 an Analytically Soluble Example} 
\author { Daniel Segel\cite{daniel}, Victor L'vov \cite {lvov}
and Itamar Procaccia\cite{procaccia}}
\address{Department of~~Chemical Physics, The Weizmann Institute of
Science, Rehovot 76100, Israel}
 \maketitle
\begin{abstract}
In turbulent flows the $n$'th  order structure functions $S_n(R)$
scale like $R^{\zeta_n}$ when $R$ is in the "inertial range". Extended
Self-Similarity refers to the substantial increase in the range of
power law behaviour of $S_n(R)$ when they are plotted as a function of
$S_2(R)$ or $S_3(R)$.  In this Letter we demonstrate this phenomenon
analytically in the context of the ``multiscaling" turbulent advection
of a passive scalar.  This model gives rise to a series of
differential equations for the structure functions $S_n(R)$ which can
be solved and shown to exhibit extended self similarity. The
phenomenon is understood by comparing the equations for $S_n(R)$ to
those for $S_n(S_2)$.
\end{abstract}
%%%%%%%%%%%%%%%%%%%%%%%%%%%%%%%
The fundamental theory of turbulence concerns itself with the
universality of the small scale structure of turbulent flows and their
characterization by universal scaling laws \cite{Fri}.  Since the days
of Kolmogorov \cite{41Kol} one expects the structure functions
$S_n(R)$ to scale with $R$ as long as $R$ is in the inertial range
\begin{equation}
S_n(R) \sim R^{\zeta_n} \,,\qquad  {\rm for~~~} \eta\ll R \ll L\ .
\label{scale}
\end{equation}
Here $\eta$ and $L$ are the viscous and outer lenghts respectively.
For a given turbulent field $u({\bf r},t)$ the $n$'th  order structure
function is defined as $S_n(R)=\left<[u({\bf r}+{\bf R},t)-u({\bf
r},t)]^n\right>$ where the angular brackets $\left<\dots\right>$
denote an average over $\bf r$ and $t$.  Komogorov's 1941 picture of
turbulence (K41) predicted that the exponents of the structure
functions of the velocity field satisfy $\zeta_n=n/3$. Further
research along the same lines suggested that the structure functions
of passive scalars densities also have the same exponents
\cite{49Obu,49Cor}.  On the other hand experiments \cite{84AGHA,87MS}
have indicated deviations from this scaling and even the possibility
of multiscaling, i.e. a non-linear dependence of $\zeta_n$ on
$n$. These findings raise fundamental questions as to our
understanding of the universality of the small scale structure of
turbulence.
 
Unfortunately, the experimental measurement of the scaling exponents
in turbulent flows is hampered by the fact that the range of scales
that exhibit scaling behaviour is rather small for Reynolds numbers Re
accessible to date. Even though the viscous scale $\eta$ decreases
like $Re^{-3/4}$, the scaling behaviour (\ref{scale}) is expected to
set in only at about $10\eta$, and cease at about $L/100$. Thus even
at $Re \sim 10^7$ the inertial range is not broad enough to allow
accurate measurements of $\zeta_n$, particularly for larger values of
$n$.

A partial resolution of these difficulties has been recently offered
 \cite{93BCTBM} by changing the way that the experimental data is
 presented. Instead of plotting log-log plots of $S_n(R)$ vs. $R$, it
 was argued that plotting log-log plots of $S_n(R)$ vs. $S_2(R)$ or
 $S_3(R)$ reveals much longer scaling ranges that can be used to fit
 accurate exponents. This method was dubbed Extended Self Similarity.
 While the effectiveness of this method has been demonstrated
 \cite{93BCBRCT}, the reason for its success remains
 unclear. Therefore its reliability is not guaranteed.  The aim of
 this Letter is to address this method theoretically for Kraichan's
 model of a passive scalar advected by a very rapidly varying velocity
 field \cite{68Kra}. In this model multiscaling is known to occur
 \cite{94Kra,95FGLP,95LP}, and we have sufficent analytic
 understanding of the mechanism of (multi)scaling to assess why
 Extended Self Similarity could hold. One should stress that in the
 case of normal scaling Extended Self Similarity can be rationalized
 rather straightforwardly \cite{94LP}. To our knowledge no analytic
 derivation of this effect is available in a case exhibiting
 multiscaling.

In this model the equation for the passive scalar concentration
$T({\bf x})$ is given by
\begin{equation}
\frac{\partial T}{\partial t} + ({\bf u \cdot \nabla})T
 = \kappa \nabla^2 T
\ , \label{passive}
\end{equation}
where the velocity field ${\bf u}$ is taken as Gaussian random and
with a $\delta$-function correlation in time.  The structure
functions $S_{2n}(R)$ are know to obey  differential
equations \cite{68Kra,95FGLP}
\begin{equation}
 R^{1-d}{\partial \over \partial R} R^{d-1}h(R)
        {\partial \over \partial R} S_{2n}(R)=J_{2n} (R)\ .
\label{balance}
\end{equation}
Here $h(R)= H (R/{\cal L} )^{\zeta_h}$ for $R<{\cal L}$ where $H$ is
a dimentional constant and $\cal L$ is  some characteristic outer
scale of the driving velocity field. The Scaling exponent $\zeta_h$
is a chosen parameter in this model, taking any desired value in the
interval $(0,2)$. The function $J_{2n} (R)$ is known exactly for $R$ in
the inertial range \cite{95FGLP}, but for our purposes we need to know
it also in the near viscous range. A form that is appropriate in both
ranges has been suggested by Kraichnan \cite{94Kra}:
\begin{equation}
J_{2n} (R)=2n \kappa { S_{2n}(R)\over S_2(R)}
\lbrack \nabla^2 S_2(R) - \nabla^2  S_2(0) \rbrack \ . \label{J2n}
\end{equation}
This form has been recently supported by numerical simulations
\cite{95Kra}.  In the inertial range the last two equations have a
multiscaling solution for $S_{2n}(R)$ with the exponents
\begin{equation}
\zeta_{2n} = \case{1}{2}\left[ \sqrt{(3-\zeta_2)^2+12 n \zeta_2} 
+\zeta_2 -3\right]\  . \
\label{zeta_2n}
\end{equation}
Our first step in studying the issue of Extended Self-Similarity using
Eqs.  (\ref{balance},\ref{J2n}) is to non-dimensionalize them. Rescale
lengths by $r_d\equiv {\cal L}(\kappa/H)^{1/\zeta_h}$ and the scalar
density by $ \Theta = r_d \sqrt{\epsilon/\kappa}$. The dimensionless
functions, $D_{2n}(y)$ are:
\begin{equation}
D_{2n}(y)\equiv   S_{2n}(y r_d) /\Theta^{2n}\,,
\qquad y\equiv R/r_d \ .
\label{def1S2n}
\end{equation}
They satisfy the equations
\begin{equation}
{d^2 D_{2n}\over dy^2} + {2+\zeta_{\eta}\over y}
 {d D_{2n}\over dy}
  ={2n D_{2n}\over D_2 y^{\zeta_{\eta}}}
~\left[ 1 -  {d^2D_2 \over d^2 y}\right] \ . \label{scaleq}
\end{equation}
The success of extended self similarity for this model can be assessed
 from Fig.1 which compares plots of $D_{2n}$ vs. $y$ with plots
 of $D_{2n}$ vs. $D_2$. The data is based on the
 numerical solution of Eq.(\ref{scaleq}) using the IMSL/IDL's
 Adams-Gear ODE-solver. The efficacy of Extended Self Similarity
 speaks for itself.

Next we exhibit the phenomenon analytically by finding the first order
dissipative correction to the inertial range scaling of the structure
functions. In the inertial range the exponent of $D_2(y)$ is $\zeta_2
= 2 - \zeta_{\eta}$ \cite{68Kra}. We can find the first correction to
the power law, going into the dissipation range, by writing $D_2(y) =2
y^{\zeta_2} /3\zeta_2 + \epsilon(y)$. We then find the following ODE
for $\epsilon(y)$:
\begin{equation}
{d^2 \epsilon\over d^2y}+ {2+\zeta_{\eta}\over  y}{d \epsilon \over dy}
+\frac{4(1-\zeta_{\eta})}{3 y^{2\zeta_{\eta}}} = 0 \ . \label{eps}
\end{equation}
There are two possibilities for the leading order solution of this
equation. If all the terms are of the same order the solution is
$\epsilon_1 = -2 y^{2-2\zeta_{\eta}}/3(1+\zeta_2)$.  If the first two
terms are larger we have $\epsilon_2 = B_2 +\epsilon_1$, where now
$\epsilon_1$ appears as the next order correction after a constant
$B_2$. One can see that the dominant solution for small $\zeta_h$ is
$\epsilon_1$, with a transition to $\epsilon_2$ at $\zeta_h=1$. Thus
the near-inertial solution for $D_2$ is
\begin{eqnarray}
 D_2(y) = \frac{2 y^{\zeta_2}}{3 \zeta_2}
\left(1 -{ \zeta_2 y^{\zeta_2-2}\over 1+\zeta_2}\right)
                         &{\rm ~~for~~}& \zeta_{\eta} < 1\,,  \\
\label{S2inert1}
 D_2(y) =  \frac{2 y^{\zeta_2}}{3 \zeta_2}
\left (1 + {C_2\over  y^{\zeta_2}}\right )
                        &{\rm ~~for~~}& \zeta_{\eta} > 1\ ,
\label{S2inert}
\end{eqnarray}
where $C_2$ is a constant. 
Note that for the case $\zeta_h=2/3$ the ODE for $D_2(y)$ can
be solved exactly in terms of elementary operations and the results
are consistent with this solution.

Next we repeat this procedure for $D_{2n}(y)$.  We substitute
 $D_{2n}(y) = A_{2n} y^{\zeta_{2n}} \lbrack 1 + \xi(y)
 \rbrack$ in Eq.(\ref{scaleq}). For the case $\zeta_h >1$ we find
\begin{equation}
{d^2\xi \over dy^2} + {2 \over y}{d \xi \over dy} (2 \zeta_{2n} 
+ 4 - \zeta_2)
+ {3n C_2 \zeta_2 \over y^{2+\zeta_2}} = 0 \ , \label{epsn}
\end{equation}
with a leading order scaling solution $\xi(y)=B_{2n} y^{-\zeta_2}$. The
constant is found from Eq.(\ref{epsn}) and the final result for $
\zeta_2 < 1 $ is
\begin{equation}
D_{2n}(y) = A_{2n} y^{\zeta_{2n} }\left[1
 + \frac{3nC_2}{(2\zeta_{2n}+3-2\zeta_2)y^{\zeta_2}}\right]
\ . \label{S2n}
\end{equation}
At this point we can assess the efficacy of Extended Self Similarity
for the case $\zeta_h >1$. If all functions $D_{2n}$ were an
exact power law in argument $D_2$ we would have
\begin{equation}
D_{2n}^{^{\rm ESS}}(y) =\left[ {2( y^{\zeta_2} + C_2 ) \over
3 \zeta_2} \right]^{\zeta_{2n}/\zeta_2} \ , \label{exactess}
\end{equation}
or in the same order
\begin{equation}
D_{2n}^{^{\rm ESS}}(y) = A_{2n} y^{\zeta_{2n}}
\left(1 + \frac{\zeta_{2n} C_2}{\zeta_2 y^{\zeta_2}}\right)
\ . \label{S2nES}
\end{equation}
The ratio $r_1$ between the coefficients of $y^{-\zeta_2}$ in Eqs.
(\ref{S2n}) and (\ref{S2nES}) measures the effectiveness of this
presentation. Using  Eq. (\ref{zeta_2n}) one finds
\begin{equation}
r_1 =
\frac{3n\zeta_2}{\zeta_{2n}(2\zeta_{2n}+3-2\zeta_2)}
= \frac{n \zeta_2}{2 n \zeta_2 - \zeta_{2n}} \ . \label{r1}
\end{equation}
  If $r_1=1$ we would have perfect Extended Self Similarity; $r_1=0$
is what we would get if we just plotted $S_{2n}$ as a function of $r$.
In Fig. 2(a) we can see a graph of $r_1$ as a function of $\zeta_2$
for $n=2, n=4$ and $n=6$. We can see that $r$ is always in the
interval $[1/2,1]$:  from Eq. (\ref{zeta_2n}) 
$\lim_{n \rightarrow \infty} r_1 = \frac{1}{2}$, and from 
Eqs. (\ref{zeta_2n}),(\ref{r1}) $\lim_{\zeta_2 \rightarrow 0} r_1
= 1$.  The convergence to $1/2$ is rather slow and not
uniform in $\zeta_2$.

The main point of the Extended Self Similarity is that $S_2$ (which is
a function of $R$) is a ``better" variable than the distance $R$
itself when it comes to scaling behaviour. Thus, to complete our
analysis we want to write the equations for $S_{2n}$ as ODE's with
respect to argument $S_2$ with the inertial range scaling plus a
remainder term. Changing variables and writing the $S_{2n}$ as a
function of $S_2$, we expect to find that the remainder term,
evaluated in terms of the inertial range estimates, is smaller than
when using $r$ as a variable. To this aim we introduce a new set of
function $F _{2n}(x)$ of argument $x=D_2(y)$:
\begin{equation}
F _{2n}(x)\equiv D_{2n}(y), \quad x \equiv
D_2(y)\ .
\label{tildeS2n}
\end{equation} 
Then Eqs. (\ref{scaleq}) gives for $F_{2n} (x)$ the following ODE
\begin{eqnarray}
{d^2 F _{2n}\over d^2 x} 
&+&{2\over 2+  [f(x)]^{2-\zeta_2}}
\left[{d F_{2n} \over d x}
-{n F _{2n}(x)\over x} \right]
\nonumber\\
&\times&
\left[ \left({d f\over dx}\right)^2 
+ {(4-\zeta_2)\over f(x) }{ d f\over dx}\right]
 = 0 \ ,
 \label{hat}
\end{eqnarray}
where $f(x)$ is the inverse function of $S_2(y)$:
$f(x)= y$ if $x=S_2(y)$.  Now let us introduce a third set of
dimensionless structure functions, $G_{2n} (z)$ such that
\begin{equation}
G _{2n}(x)\equiv D_{2n}(y), \quad x \equiv
2y^{\zeta_2/3\zeta_2}\ .
\label{def2S2n}
\end{equation}
Note that the functions $F_{2n}$ and $G_{2n}$ are identical in the
inertial range, since one is a function of $S_2$ and the other is a
function of its power law in the inertial range. Indeed, it will
be seen below that they satisfy the same ODE in the inertial range
with a power solution $x^{\zeta_{2n}/\zeta_2}$.  Using (\ref{scaleq})
one derives the following ODE for $G_{2n}(x)$:
\begin{eqnarray}
{d^2 G_{2n} \over dx^2}&+& 
{3 \over \zeta_2 x}\left[ {d G_{2n} \over dx}
-\frac{ n  G_{2n} (x)}{ g(x)}\right] \nonumber \\
&+& \frac{3 n  G_{2n} (x) }{g (x) x^{2/\zeta_2}}
\left[ (\zeta_2-1){dg \over dx}
+ \zeta_2 x{d^2 g \over dx^2 }\right]=0
\label{check}
\end{eqnarray}
with the function $g(x)$ satisfying $g(x=D_2(y))=D_2(y)$.

Consider first the $\zeta_2 < 1$ case, and use the near-inertial
approximation for $S_2(y)$, Eq. (\ref{S2inert}).  Inverting this
relation we find the relevant approximation for $f$ in terms of $x$,
i.e.
\begin{equation}
f(x) = \left(\frac{3\zeta_2\, x}{2}\right)^{1/\zeta_2}
\left(1 - \frac{2 C_2}{3 \zeta_2^2\, x} \right) \ . \label{fx}
\end{equation}
Using this and introducing constants $ C_{2n}^0$ $S_{2n} = C_{2n}^0
x^{\zeta_{2n}/\zeta_2}$ in the remainder term we find that the
near-inertial version of Eq. (\ref{hat}) is
\begin{eqnarray}
{d^2 F_{2n} \over d^2 x}
&+& {3 \over \zeta_2 x}\left[ {d F_{2n}   \over dx}
-\frac{ n  F_{2n} (x)}{x}\right]
\label{near}\\
 &+& ~\frac{2 C_2 C_{2n}^0}{\zeta_2^3}(\zeta_{2n}-n\zeta_2)
x^{\zeta_{2n}/\zeta_2-3}= 0 \ . 
\nonumber
\end{eqnarray}
Doing the same to Eq. (\ref{check}) we have
\begin{eqnarray}
{D^2 G_{2n}\over d^2 x}
&+& {3 \over \zeta_2 x}\left[ {d G_{2n}\over dx}
-\frac{ n  G_{2n}(x)}{x}\right]
\label{s''}\\
 &+& ~\frac{2 C_2 C_{2n}^0}{\zeta_2^3}
n\, x^{\zeta_{2n}/\zeta_2-3}= 0 \ .
\nonumber
\end{eqnarray}
Indeed, we see that the first lines of these equations , which are the
inertial range contributions, are the same. 
Also the second lines are of the same order in $y$ but the ratio $r_2$
between the coefficients in these lines (reminder terms) is
\begin{equation}
r_2 = \frac{\zeta_{2n} - n\zeta_2}{n \zeta_2} \ . \label{r2}
\end{equation}
Since the H\"older inequalities imply that $\zeta_{2n} \leq n\zeta_2$,
 we have $|r_2| \leq 1$; this means that indeed the remainder term
 that ``spoils" the inertial range scaling to first order is smaller
 in the new representation.  One can see, by solving Eq. (\ref{hat})
 in terms of the coefficient of the remainder term, using $F_{2n} (x)$
 to express $D_{2n}(y)$ back again and comparing with Eq. (\ref{S2n}),
 that one should have the relation
\begin{equation}
r_2 = 1 - \frac{1}{r_1} \ . \label{r22}
\end{equation}
Now comparing Eqs. (\ref{r1}) and (\ref{r2}) one sees that this
relation is true.  Extended Self Similarity as represented by $r_1$
very near unity indeed translates into very small $r_2$, that is into
a small remainder term.  In Fig. 2(b) we can see $r_2$ as a function
of $\zeta_2$ for $n=2, n=4$ and $n=6$. As $n \to \infty$ the ratio
becomes poor, $r_2 \to -1$ but the convergence is slow as we can see
from Fig. 2(b) and again not uniform in $\zeta_2$, since one can
verify that $\lim_{\zeta_2 \rightarrow 0} r_2 = 0$.

The situation for $\zeta_2 > 1$ is much messier algebraically although
just the same procedure can be followed in principle.  We have already
given in Eq. (\ref{S2inert}) the near-inertial form for $D_2(y)$ for
$\zeta_2 > 1$; we now substitute this and $D_{2n}(y) = A_{2n}
y^{\zeta_{2n}} \lbrack 1 + \xi(y) \rbrack$ [considered as a definition
of $\xi(y)$] into Eq. (\ref{scaleq}) to find the equation
\begin{eqnarray}
{d^2 \xi\over dy^2}&+&(2 \zeta_{2n} + 4 - \zeta_2) {d \xi\over y dy}
 \nonumber \\
&+& {n \zeta_2 (1+2\zeta_2)(2-\zeta_2)\over (\zeta_2+1)}
 y^{\zeta_2-4}  = 0 \ . \label{xi''}
\end{eqnarray}
This equation has an obvious power law solution which leads to 
\begin{equation}
D_{2n}(y) = A_{2n} y^{\zeta_{2n}}\left[
1 + \frac{n \zeta_2(1+2\zeta_2)}{(\zeta_2+1)(1+2\zeta_{2n})}
y^{\zeta_2-2}\right]
\label{S2ny}
\end{equation}
(for $\zeta_2 > 1)$, which gives
\begin{equation}
r_1 = \frac{n\zeta_2
 (1+2\zeta_2)}{\zeta_{2n}(1+2\zeta_{2n})} \ . \label{r12}
\end{equation}
In Fig. 3 we can see $r_1$ as a function of $\zeta_2$ for $n=2, n=4$
and $n=6$.  For very large $n$ we can see that
\begin{equation}
\lim_{n \rightarrow \infty} r_1 = \frac{1+2\zeta_2}{6} \in
[\frac{1}{2},\frac{5}{6}] \ , \label{lim}
\end{equation}
but the convergence is slow as we can see from Fig. 3 and again not
uniform in $\zeta_2$, since one can verify that $\lim_{\zeta_2
\rightarrow 0} r_1 = 1$.  Again one can verify that the efficacy of
ESS is reflected in a smaller remainder term in the ODE for $S_{2n}$
using $S_2$ as a variable, as expressed by a ratio $r_2$ for which
here too $r_2 = 1 - r_1^{-1}$.

The success of the method in this case cannot be glibly extended to
any turbulent system.  We see from the analysis that the improvement
in the scaling behaviour of $S_n$ when presented as a function of
$S_2$ is {\it not} a fundamental result but a statement on the similar
behaviour of the structure functions going into the dissipation
range. This is measured by the value of the parameter $r_1$ of
Eq.(\ref{r1}). We can imagine in principle - although no examples are
known - a system for which the dissipative scale for $S_n$ is {\it
smaller} than that for $S_2$. In such a system the method will not
improve the scaling plots.  In this passive scalar model the
dissipative scales of all the structure functions are of the same
order, as has been shown in \cite{95LP}. This is one of the deep
reasons for the success of the method in the present case. It is
tempting to conjecture that the success of the method in Navier-Stokes
turbulence is an indication that the dissipative scales of $S_n$ are
either of the same order or become larger with $n$. The full
understanding of the method in that case must await a better analytic
understanding of the structure functions and their functional form in
the near dissipative regime.

Figure Legends:

Fig.1: The dimensionless structre functions $D_{2n}(y)$ plotted in log-log
plots as functions of $y$ (solid line) and $S_2(y)$ (dashed-dotted line)
respectively. In dashed line the exact scaling law is shown.
In panels (a) and (b)
$\zeta_h=1/2$ and $n=2$ and $n=6$ respectively. Panels (c) and (d) show
results for $\zeta_h=3/2$ and the same value so of $n$. One sees an 
improvement of the scaling law in at least 54 orders of magnitude ???.

Fig.2: panel (a): The ratio $r_1$ of Eq.(15) as a function of $\zeta_2$
for $n=2$ (solid line), $n=4$ (dashed line) and $n=6$ (dotted-dashed line).
panel (b): The ratio $r_2$ of Eq.(25) for the same values of $n$. 

Fig.3 The ratio $r_1$ of Eq.(15) for $\zeta_2>1$, and the same values of
$n$ as in Fig.2.
\end{document}